\documentclass[twocolumn,superscriptaddress,amsmath,amssymb,aps,physrev]{revtex4-1}

\usepackage{graphicx}
\usepackage{dcolumn}
\usepackage{bm}
\usepackage{siunitx}
\usepackage{upgreek}
\usepackage{graphicx}
\usepackage[usenames,dvipsnames]{xcolor}

\usepackage[colorlinks=true]{hyperref}  
\hypersetup{
    bookmarks=true,         
    unicode=false,          
    pdftoolbar=true,        
    pdfmenubar=true,        
    pdffitwindow=false,     
    pdfstartview={FitH},    
    pdftitle={vortex_dynamics},    
    pdfauthor={Perego},     
    pdfsubject={},   
    pdfcreator={},   
    pdfproducer={}, 
    pdfkeywords={} {} {}, 
    pdfnewwindow=true,      
    colorlinks=true,       
    linkcolor=blue, 
    citecolor=blue,        
    filecolor=magenta,      
    urlcolor=blue           
}

\begin{document}

\title{\textbf{Vortex Dynamics in Magic-Angle Twisted Graphene}} 
\author{Marta Perego}	
\email{mperego@phys.ethz.ch}
\author{Peter Koopmann}
\author{Clara Galante Agero}
\author{Alexandra Mestre Tor\`a}
\affiliation{Laboratory for Solid State Physics, ETH Zurich,~CH-8093~Zurich, Switzerland}
\affiliation{Quantum Center, ETH Zurich,~CH-8093 Zurich, Switzerland}
\author{Takashi Taniguchi}
\affiliation{Research Center for Materials Nanoarchitectonics, National Institute for Materials Science,  1-1 Namiki, Tsukuba 305-0044, Japan}
\author{Kenji Watanabe}
\affiliation{Research Center for Electronic and Optical Materials, National Institute for Materials Science, 1-1 Namiki, Tsukuba 305-0044, Japan}
\author{Vadim Geshkenbein}
\affiliation{Institute for Theoretical Physics, ETH Zurich,~CH-8093~Zurich, Switzerland}
\author{Gianni Blatter}
\affiliation{Institute for Theoretical Physics, ETH Zurich,~CH-8093~Zurich, Switzerland}
\affiliation{Quantum Center, ETH Zurich,~CH-8093 Zurich, Switzerland}
\author{Thomas Ihn}
\author{Klaus Ensslin}
\affiliation{Laboratory for Solid State Physics, ETH Zurich,~CH-8093~Zurich, Switzerland}
\affiliation{Quantum Center, ETH Zurich,~CH-8093 Zurich, Switzerland}

\date{\today}

\begin{abstract}
We use a gate-defined Josephson junction (JJ) device made from twisted-layer
graphene for studying vortex dynamics in two dimensions.  The JJ sensor
signals the presence of individual vortices in the superconducting leads
nearby the junction through shifts in the Fraunhofer interference pattern of
the magnetic-field-dependent critical current $I_c(B)$ across the junction.
Rapid vortex fluctuations manifest as telegraph-type noise in time traces of
the junction voltage $V(t)$. Measurements of $I_c(B)$ and $V(t)$ are
interpreted in terms of multi-vortex processes where fast vortex fluctuations
in the leads are modulated by quasi-stationary vortices trapped in the leads.
The different timescales associated with these processes allow for their
disentangling and quantitative analysis.  Tracking the temperature dependence
of the vortex-dynamical rates between $T = \SI{7}{mK}$ and $T = \SI{120}{mK}$,
we find that the creep type vortex motion is thermally activated above $T
\approx \SI{100}{mK}$, while the saturation of rates below $T \approx
\SI{80}{mK}$ is suggestive of a sharp transition to macroscopic quantum
tunneling of vortices.
\end{abstract}

\maketitle

\section{Introduction} \label{sec:Int}
Twisted graphene films have emerged as novel superconductors with remarkable
tunability, enabling manipulation of the superconducting state through a mere
change in carrier density \cite{cao2018correlated, yankowitz2019tuning,
lu2019superconductors, cao2018unconventional, park2021tunable,
hao2021electric, carr2017twistronics, park2022robust,kim2022evidence,
park2022robust, zhang2022promotion, burg2022emergence,
mukherjee2025superconducting}. Local gates allow for tuning different regions
of the same sample into distinct phases, superconducting, metallic, or
insulating; using such technique, the homogeneous material can be structured
into a functional device \cite{Rodan-Legrain2021, deVries2021,
diez2023symmetry, portoles2022tunable, diez2025probing, Ronen2025,
wakiri2024tunable, zheng2024gate}. Numerous applications in superconducting
electronics have emerged by now \cite{deVries2021, Rodan-Legrain2021,
portoles2022tunable, diez2023symmetry, rothstein2025gate, zheng2024gate,
wakiri2024tunable, jha2025large, diez2025probing, Ronen2025}; here, we report
on a device made from twisted-layer graphene that allows for the study of the
material's superconducting phenomenology as determined by vortex physics
\cite{Blatter1994}.

Superconducting structures define inductive devices governed by magnetic
fields and currents rather than electric fields and charges. In type II
superconductors, topological defects, i.e., vortices, determine the material's
response to electrical currents and magnetic fields
\cite{Abrikosov1957,Tinkham2004}.  These vortices and their dynamics are
usually measured as bulk ensembles, with notable exceptions such as
single-vortex imaging through magneto-optical imaging \cite{Bending1999},
scanning tunneling- \cite{Hess1990} and Lorentz microscopy
\cite{Tonomura2001}, or the SQUID on Tip technique \cite{Embon2015}.
Bulk ensemble measurements of magnetic moments and their time
decay inform about vortex penetration and their creep dynamics
\cite{Anderson1962, AndersonKim1964, Beasley1969} in a material. The topic
gained much interest with the cuprate high-$T_c$ materials
\cite{YeshurunMalo1988, Campbell1990}, where fluctuations are particularly
relevant and glassy vortex dynamics has been uncovered \cite{Maley1990,
Yeshurun1996}. The saturation in the decay of magnetization to a finite value
at low temperatures \cite{mota1992quantum, mota1994flux, fruchter1991low,
hoekstra1999temperature} was attributed to quantum creep \cite{glazman1984,
Blatter1991}, the macroscopic quantum tunneling of vortices. Besides magnetic
relaxation, the analysis of the voltage--current $V$--$I$ characteristics
below critical field or temperature provides information on
vortex creep \cite{Buchacek2019}, e.g., the quantum dynamics of Pearl-vortices
\cite{Pearl1964} in high-$T_\mathrm{c}$ thin films \cite{tafuri2006}.

Here, we take a next step in the observation of vortex creep by studying the
dynamics of individual vortices. To do so, we use a Josephson junction (JJ) in
a thin-film device made of twisted-layer graphene as a vortex sensor
\cite{perego2024experimental} enabling detection of individual Pearl-vortices
\cite{Pearl1964} penetrating or leaving the superconducting leads. Our study
adds a new transport-based probe for observing individual vortex penetration
and dynamics, yielding rates of vortex motion as a function of experimental
parameters such as carrier density, temperature, or magnetic field. Our
findings show a complex vortex dynamical behavior involving different time
scales, with vortices penetrating a twisted-layer graphene film by thermal
activation over barriers at high temperatures and by quantum tunneling through
barriers at low temperatures.

In our previous work using the single-vortex Josephson sensor, we have
identified single-vortex entry- and exit-processes in superconducting leads
through the observation of abrupt jumps in the Fraunhofer-type interference
pattern of the Josephson junction \cite{perego2024experimental}; furthermore,
the observation of vortex fluctuations at high temperatures has been used to
infer the film's superfluid density $\rho_s$. In
Ref.~\cite{perego2026tunneling}, we have reported on the single-vortex
dynamics manifested through the observation of telegraph-type noise in the
junction voltage. The temperature dependence of the measured vortex-dynamical
rates, in particular their saturation to a finite value at low temperature,
has led us to propose the observation of macroscopic quantum tunneling of
vortices in this material. In the present paper, we analyze the complex
dynamics of vortices uncovered by our single-vortex detector.

In the following, see Sec.~\ref{sec:JJ}, we discuss the setup 
and functionality of our Josephson-junction vortex-sensor. In section
\ref{sec:FP}, we describe the noise visible in the Fraunhofer pattern that we
attribute to rapid vortex fluctuations in the leads ((sub-)seconds timescale).
Section \ref{sec:TN} is devoted to the analysis of the voltage traces $V_\mathrm{d}(t)$
associated with the noise previously seen in the Fraunhofer pattern. We
identify two types of segments in the traces with separated time scales that
we associate with the presence/absence of an additional distant
quasi-stationary vortex (timescale of order $\SI{e3}{s}$). We extract the
rates associated with the various processes and present our interpretation
thereof in Sec.~\ref{sec:Dis}, see also our previous publications
\cite{perego2024experimental, perego2026tunneling}. We summarize our findings
in Sec.~\ref{sec:Sum}.

\section{Josephson junction as single-vortex sensor}\label{sec:JJ}

\begin{figure}
\includegraphics[width=\columnwidth]{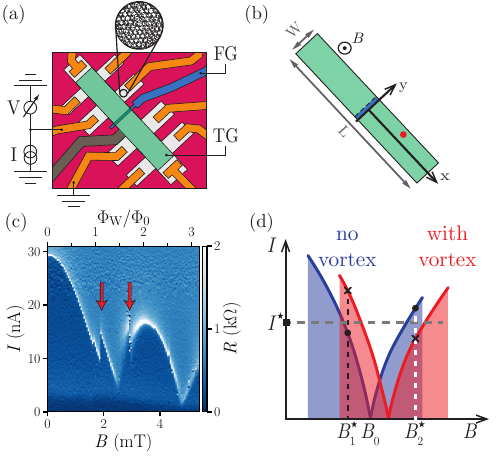}
\caption{(a) Schematic of twisted-layer graphene device with contacts
(orange), top gate (green, TG) and finger gate (blue, FG) defining the
Josephson junction (the graphite bottom gate is below the entire structure and
not shown here). (b) Device geometry with width $W = \SI{1.1}{\upmu m}$ along
the $y$-direction, length $L = 6W$ along $x$, and the junction width $L_{\rm
j} = \SI{150}{nm}$. The film thicknness is $d = \SI{1}{nm}$. An applied
magnetic field $H$ perpendicular to the plane penetrates the weakly screening
sample (with the Pearl length $\Lambda = 2 \lambda_{\rm\scriptscriptstyle
L}^2/d$ much larger than the film width $W$, $\lambda_{\rm\scriptscriptstyle
L}$ is the London length) such that $B \approx H$.  (c) Vortices (red dot in
(b)) penetrating the sample nearby the junction are detected as jumps in the
Fraunhofer-type pattern, see red arrows (result for the {\it strong-leads}
sample at $T=\SI{7}{mK}$). The top axis shows the corresponding scaled flux $\Phi_\mathrm{W}/\Phi_0$ penetrating junction and leads, where $\Phi_\mathrm{W} = BW^2$. (d) Schematic illustrating the change in the
dissipation upon vortex entry into a lead: Measuring the voltage $V$ across
the junction at fixed current $I^\ast$ and field $B_1^\ast < B_0$
(black-dashed line), with $B_0$ the zero of the FP, the response is
dissipative when no vortex is present in the lead ($I_{c}^{\rm nv} < I^\ast$),
while the junction is superconducting in the presence of a vortex ($I_{c}^{\rm
v} > I^\ast$) and $V = 0$. When the measurement field $B_2^\ast > B_0$
(white-dashed line) resides above the zero of the FP, the response is
reverse.}
\label{Fig1} 
\end{figure}

Our single-vortex detector is a gate-defined Josephson junction (JJ) made of
magic-angle twisted four-layer graphene (MAT4G), previously studied in
Refs.~\cite{perego2024experimental,perego2026tunneling}. Gating and geometry
of the device are shown in Fig.~\ref{Fig1}(a) and (b); the film width $W$ is
much smaller than the length $L$ ($W\ll L$).  Choosing appropriate gate
voltages, we define two tunings: for the {\it strong-leads} configuration
\cite{perego2024experimental}, the lead density $n_\textnormal{l} =
\SI{4.2e12} {cm^{-2}}$ is tuned to the middle of the electron superconducting
dome, where the critical current and the critical temperature take the highest
values, while for the {\it weak-leads} configuration
\cite{perego2026tunneling} the lead density is tuned to the edge of the
electron superconducting dome with $n_\textnormal{l} = \SI{4.8e12}{cm^{-2}}$.
The carrier density in the junction region is tuned to the resistive state at
full filling in both cases (see Appendix \ref{appendix_A} for tuning details).

In our device, vortex detection is predicated on their influence on the
Josephson phase of the junction \cite{Golod2010}. Upon application of a
perpendicular magnetic field $B$, the junction critical current $I_c(B)$
exhibits a Fraunhofer-like interference pattern (FP) \cite{Tinkham2004,
Clem2010}, an example of which is shown in Fig.~\ref{Fig1}(c) ({\it
strong-leads} tuning at $T = \SI{7}{mK}$).  The color in this figure encodes
the differential resistance $R = d V/d I$, with dark blue describing the
superconducting state ($R = 0$) below $I_c(B)$ and lighter blue associated
with the resistive state above.  The white trace separating the two regions
testifies for the sharp rise in the $V-I$ characteristic at the critical
current $I_c(B)$. As compared to the standard FP \cite{Tinkham2004} with
maxima decreasing as $1/B$, under the present weak-screening conditions with
$\Lambda = 2 \lambda^2_{\rm\scriptscriptstyle L}/d \gg W$ ($\Lambda$ and
$\lambda_{\rm\scriptscriptstyle L}$ denote the Pearl- \cite{Pearl1964} and
London lengths), the maxima decrease more slowly, as $1/\sqrt{B}$;
furthermore, the field-to-flux transformation involves the area $W^2$
\cite{Clem2010} rather than the usual area $(2\lambda_{\rm \scriptscriptstyle
L} + L_{\rm j})\,W$ \cite{Tinkham2004}. The impact of vortex entry or exit on
the FP is highlighted by the red arrows in Fig.~\ref{Fig1}(c). For $B>0$, the
entry of a vortex into the superconducting leads {\it decreases} the effect of
the magnetic field $B$ on the Josephson phase, resulting in a shift of the FP
towards higher fields (to the right) \cite{Golod2010,Clem2011, KoganMints2014,
KoganMints_PC2014}, as schematically illustrated in Fig.~\ref{Fig1}(d).
Conversely, a vortex exiting the leads (at $B>0$) induces a shift of the FP
towards lower fields $B$ (to the left). The direction of these vortex-induced
shifts can be easily understood when comparing the Meissner- and vortex
currents: While the Meissner currents {\it shield} the outside field $H$, the
vortex currents {\it hold} the inside flux $\Phi_0$ of the vortex, hence, they
run in opposite directions. As a result, the field-induced currents
$j_y(x=0^\pm)$ at the junction (determining the phase change $\Delta
\gamma_{\rm \scriptscriptstyle J}(y)$ across the junction \cite{Clem2010}) are
reduced by the presence of a vortex in the leads.

Beyond the simple shift of the Fraunhofer pattern described above, depending
on the location $(x_{\rm v},\, y_{\rm v})$ of the vortex trapped in the lead
(red dot in Fig.~\ref{Fig1}(b)), the Fraunhofer pattern may
undergo additional distortion. The above simple picture, where the presence of
a vortex mimicks a reduction in the $B$-field, holds for vortices located at a
distance $x_{\rm v} > 0.5\,W$ from the junction (assuming $y_{\rm v} = 0$).
Vortices trapped closer to the junction, i.e., $x_{\rm v} < 0.5\,W$, produce
an increasingly steeper phase drop upon reducing the distance $x_{\rm v}$,
different in shape from the smoothly increasing field-induced phase difference
$\Delta \gamma_{\rm \scriptscriptstyle J}(y) \approx 1.7 (B W^2/\Phi_0)
\sin(\pi y/W)$. This leads to an additional distortion of the Fraunhofer
pattern on top of a shift, resulting in a right-shifted asymmetric Fraunhofer
pattern.  Finally, vortices trapped at finite distance $|y_{\rm v}| > 0$ away
from the lead middle lift the sharp zeroes in the Fraunhofer pattern. For a
more quantitative understanding of the vortex-induced changes in the
Fraunhofer pattern, we refer the reader to Figs.~14--16 in
Ref.~\cite{perego2024experimental} (Supplementary Material).

Summarizing, vortex penetration into the superconducting leads manifests as
shifts (jumps) of the Fraunhofer pattern. Vortices trapped not too close to
the junction mimick a {\it reduction} in the field magnitude and thus a
rightward shift of the FP.  The observed shifts appear as sharp events
(without internal structure). The field scale $\Delta B$ defining the
separation between zeroes in the Fraunhofer pattern is determined by the unit
flux $\Phi_0 = h/2e$ penetrating the area $W^2$ of the lead \cite{Clem2010,
perego2024experimental}, $\Delta B \approx 1.8 \Phi_0/W^2$; the observed
shifts in the FP then correspond to a change in flux measuring a fraction of
the flux quantum $\Phi_0$, see Fig.~\ref{Fig1}(c). From this observation, we
conclude that we deal with single-vortex events \cite{Golod2010,Clem2011,
KoganMints2014, KoganMints_PC2014}.

The \textit{strong-leads} tuning involves a large superfluid density $\rho_s$
and correspondingly large energy barriers $U \approx \varepsilon_0 d = \pi
\rho_s$ for vortex entry (exit) into (out of) the film as
previously demonstrated in Ref.~\cite{perego2024experimental}. Here,
$\varepsilon_0 = (4\pi/\mu_0)(\Phi_0/4\pi \lambda_{\rm \scriptscriptstyle
L})^2$ is the vortex line energy, $\mu_0$ is the vacuum permeability, and $d$
is the film thickness.  With these large energy barriers, temporal vortex
fluctuations are suppressed, with typical lifetimes on the order of
$\sim\SI{e3}{s}$. As a result, individual vortex jumps into and out of the
film are well resolved, as documented in Fig.~\ref{Fig1}(c).

In order to study vortex dynamics, we wish to reduce the time-scale of vortex
entry and exit by reducing  the creep barriers, i.e., the superfluid stiffness
$\rho_s$.  This is achieved by tuning the film to the edge of the
superconducting dome, i.e., into a {\it weak-leads} tuning regime, see
Appendix~\ref{appendix_A} for tuning details.  This opens the way for faster
vortex dynamical events on a time scale around and below $\sim\SI{1}{s}$ that
we are going to study in the following. Before doing so, let us briefly pause
and discuss an alternative scenario for potential jumps in the Fraunhofer
pattern.

\subsection{Pearl-Josephson- versus Pearl-vortices}\label{sec:AJPvortices}

An interesting question is the location of the topological defects triggering
the jumps in the Fraunhofer pattern---are these jumps due to Pearl-vortices in
the leads or rather due to metastable configurations of Josephson-vortices in
the junction? First, we have to distinguish between one period along $y$ (see
Fig.~\ref{Fig1}(b)) in the junction current density $j(y) = j_c \cos[2\pi
(\Phi(B)/\Phi_0) y/W]$, with $\Phi(B) \approx 2\lambda_{\rm \scriptscriptstyle
L} W H$, within a standard (short) Josephson junction \cite{Tinkham2004} versus
the solitonic solution of extension $\lambda_{\rm \scriptscriptstyle J} \gg W$
of the sine-Gordon equation in a long junction \cite{Tinkham2004} (with
$\lambda_{\rm \scriptscriptstyle J} = (\Phi_0/4\pi \mu_0 \lambda_{\rm
\scriptscriptstyle L} j_c)^{1/2}$ the Josephson length)---both are referred to
as `Josephson-vortices'. While the current pattern in a short junction of
length $W \lesssim \lambda_{\rm \scriptscriptstyle J}$ is pinned to the edges,
the solitonic Josephson-vortices in a long junction ($W \gg \lambda_{\rm
\scriptscriptstyle J})$ can be easily shifted along the junction, i.e., along
$y$; hence, the latter have their own dynamics and can be pinned to the edges
as well as defects along the junction, leading to (multiple) metastable
junction-states \cite{owen1967} that could produce jumps in the Fraunhofer
pattern. Thus we have to check whether we deal with a `short' or a `long'
Josephson junction.

Our junction device involves thin, weakly-screening leads and a rather
strong-coupling junction. This combination renders the structure of
(dynamical) Josephson vortices non-trivial.  The situation is similar to the
one in bulk junctions as discussed by Gurevich \cite{gurevich1992}. In
strongly coupled junctions, the Josephson vortex involves two scales: the size
of the phase-core $2\xi_{\rm \scriptscriptstyle J} = 2\xi/D \approx
\lambda_{\rm \scriptscriptstyle J}^2/\lambda_{\rm \scriptscriptstyle L}$
(where the phase changes by $\sim 2\pi$, $D$ denotes the junction transparency
$D = j_{c{\rm \scriptscriptstyle J}}/j_0$, the ratio of the junction and
depairing critical current densities), and its magnetic size $\lambda$ (where
currents vanish). In a bulk situation \cite{gurevich1992}, the so-called
Abrikosov-Josephson vortex with two scales persists until $\xi_{\rm
\scriptscriptstyle J} \sim \lambda$, with $\lambda = \lambda_{\rm
\scriptscriptstyle L}$. For our weak-screening film \cite{Kogan2001}, the
corresponding criterion involves the Pearl-length $\lambda = \Lambda =
2\lambda_{\rm \scriptscriptstyle L}^2/d$. With $\xi \approx \SI{40}{nm}$,
$\lambda_{\rm \scriptscriptstyle L} \approx \SI{2.7}{\upmu m} - \SI{2.8}{\upmu
m}$, see Refs.~\cite{perego2024experimental,perego2026tunneling}, and a
transparency $D \approx 1/40$ ($j_{c {\rm \scriptscriptstyle J}} \approx
\SI{9.1e2}{A/cm^2}$ from experiment, $j_0 = \Phi_0/3\sqrt{3}\pi
\mu_0\xi\lambda_{\rm \scriptscriptstyle L}^2 \approx \SI{3.5e4}{A/cm^2}$), we
find a phase core $2\xi_{\rm \scriptscriptstyle J} \approx \SI{3}{\upmu m} \ll
\Lambda \approx \SI{2}{cm}$.  The Pearl-Josephson vortex in a planar junction
separating weakly-screening leads then has the following structure
\cite{Kogan2001}: a phase core with the phase difference $\Delta \gamma_{\rm
\scriptscriptstyle J}(|y| < \xi_{\rm \scriptscriptstyle J}) \approx \pi + 2
\arctan(y/\xi_{\rm \scriptscriptstyle J})$ across the junction, followed by an
intermediate regime $\xi_{\rm \scriptscriptstyle J} < |y| < \Lambda$ where
corrections decrease as $2\xi_{\rm \scriptscriptstyle J}/|y|$, before entering
the asymptotic regime $\Lambda < |y|$ of a Pearl-vortex where tails disappear
as $2\Lambda\xi_{\rm \scriptscriptstyle J}/y^2$.

The relevant size for our problem is the phase core $2\xi_{\rm
\scriptscriptstyle J}\approx\SI{3}{\upmu m}$ where the phase difference
$\Delta\gamma_{\rm \scriptscriptstyle J}$ across the junction changes by
nearly $2\pi$. This phase core is larger than the width $W=\SI{1.1}{\upmu m}$ of
the junction and we conclude that our device resides in the short-junction
limit $W\leq 2\xi_{\rm \scriptscriptstyle J}$ with a static current-pattern in
the junction.  Thus we associate the observed jumps in the Fraunhofer pattern
with the dynamics of Pearl-vortices in the leads rather than metastable states
of Pearl-Josephson vortices in the junction.

\begin{figure*}
\includegraphics[width=2\columnwidth]{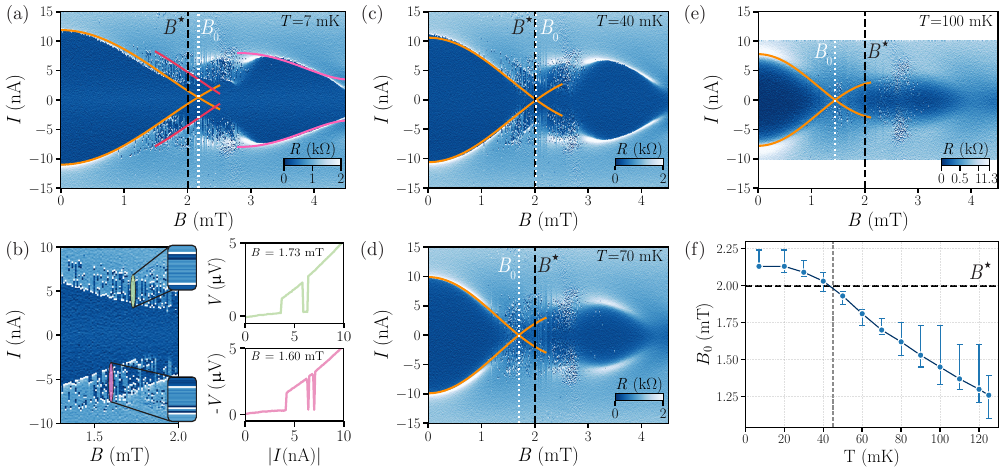}
\caption{(a) Fraunhofer interference pattern of the device with {\it weak-leads}
tuning at $T = \SI{7}{mK}$. The noisy feature around the first zero $B_0$ of
the FP is associated with fast vortex fluctuations in the leads, causing jumps
in the $V$-$I$ characteristic and thereby noise in the differential resistance
$R$.  The orange fit of the main peak allows for an estimate of $B_0$. Red
lines are a fit to the termination line of the noise, generated by a
vortex-induced shift of the FP.  Magenta lines are a fit to the second peak in
the FP with a stationary vortex very close to the junction. (b) Magnification
of the noise in the FP in (a) at two field values together with the
corresponding $V$-$I$ traces.  Superconducting (dark blue), resistive (light
blue), and jump regions (white) can be identified and correlated to the jumps
in the characteristics.  (c) -- (e) Fraunhofer patterns measured at
temperatures $T = 40,~70,~100$~{mK}.  The fits of the main peaks (orange
lines) allow to extract the first zero $B_0$ of the FP (white dotted lines) at
different temperatures.  Black dashed lines mark the field $B^\ast =
\SI{2}{mT}$ where time traces $V_\mathrm{d}(t)$ exhibiting telegraph-type noise are
measured.  (f) Summary of fits (see (a) and (c) -- (e)) of main peaks (orange
lines) to extract the temperature dependence of the first zero $B_0(T)$ of the
FP.}
\label{Fig2}
\end{figure*}

\section{Noise in Fraunhofer Pattern}\label{sec:FP}

In Fig.~\ref{Fig2}(a), we show the Fraunhofer pattern measured at the
temperature $T = \SI{7}{mK}$ for the {\it weak-leads} tuning. As the critical
current $I_c(B)$ drops to zero in the vicinity of the first zero at $B_0$ in
the Fraunhofer pattern (i.e., $I_c(B_0) = 0$), we observe a noisy pattern that
exhibits a characteristic shape reminding of a shifted FP as seen previously
in Fig.~\ref{Fig1}(c).  The magnification in Fig.~\ref{Fig2}(b) identifies
this noise as a series of white, light- and dark-blue dots resulting from
jumps between dissipative and superconducting states in the $V$--$I$ traces
(values $R < 0$ associated with downward jumps are set to zero, resulting in
dark-blue dots). Examples of these jumps are displayed in the $V-I$
characteristics, where the voltage switches between the superconducting and
the dissipative state while the bias d.c.\ current $I$ is swept at fixed field
B, see right panel in Fig.~\ref{Fig2}(b). With typical current sweeping rates
of order $\Gamma_I = \SI{0.16}{nA/s}$, such jumps appear on the time scale of
seconds.  We associate these jump-signals (involving dark- and light-blue dots
describing superconducting and resistive states separated by steep up- and
downward jumps identified as white and very-dark blue dots) with vortices
entering and exiting the leads on the time scale of few seconds.  We note that
the voltage measured in the $V$--$I$ characteristic arises from the switching
of the junction to the resistive state when the driving current $I$ surpasses
the critical junction current $I_c(B)$. The presence or absence of the vortex
in the lead then alters the critical current $I_c(B)$ via changing the
junction's phase pattern \cite{Golod2010, Clem2011, KoganMints2014,
KoganMints_PC2014}.

The overall structure of the noise, ending at a shifted slope of the
Fraunhofer pattern (see red lines in Fig.~\ref{Fig2}(a)), has an immediate
explanation: the vortices penetrating the leads follow the same path through
the leads that defines a unique maximal shift in the Fraunhofer pattern.  Such
a vortex-induced shift can be calculated, see Refs.~\cite{Clem2011,
KoganMints2014, KoganMints_PC2014} and our previous discussion in
Ref.~\cite{perego2024experimental}. The red line in Fig.~\ref{Fig2}(a) is a
fit to the data involving a vortex at the position $x_{\rm v} = 0.67\,
W,~y_{\rm v} = 0$, i.e., in the lead middle close to the junction. The vortex
changes the phase difference $\Delta \gamma_{\rm \scriptscriptstyle J}(y)$
across the junction, leading to a rightward shift of the Fraunhofer pattern by
about $\SI{0.4}{mT}$.  This interpretation is further confirmed by the
$V$--$I$ trace in Fig.~\ref{Fig2}(b) which exhibits jumps between the critical
current values $I_c$ and $I_c^{\rm v} (> I_c)$ without and with a vortex
present.  We note, however, that not all jumps visible in the $V$-$I$ traces
of Fig.~\ref{Fig2}(b) are associated with vortex entry or exit processes in or
out of the leads; specifically, the first and last upward jumps may be
associated with the junction going resistive upon surpassing the critical
currents $I_c$ and $I_c^{\rm v}$, respectively.  Furthermore, the opposite
applies as well: e.g., vortex-dynamical processes that preserve the
superconducting or dissipative state do not produce a jump in the $V$--$I$
trace and thus are not trivially detected by our vortex sensor.

While we interpret the noisy shift of the Fraunhofer pattern with rapid vortex
fluctuations in the leads, we associate the second maximum around $B \approx
\SI{3.3}{mT}$ with a quasi-stationary vortex trapped very close to the
junction.  Specifically, the fitted magenta lines in Fig.~\ref{Fig2}(a)
originate from a calculation accounting for a vortex trapped at $x_{\rm v} =
0.055\, W,~y_{\rm v} = 0.12\, W$. The presence of such a vortex reduces the
main peak at $B = 0$ and enhances the second peak in the Fraunhofer pattern,
resulting in the magenta lines shown in Fig.~\ref{Fig2}(a).  Admittedly, an
alternative interpretation of this second peak could be given in terms of a
non-uniform distribution of the critical current density $j_c(y)$ along the
junction with maxima near the junction edges and no vortices present
\cite{hart2014induced}.  However, such an interpretation is less suitable to
fit the main peak of the Fraunhofer pattern and we do not follow this
possibility further on.

The association of lead-states, pristine or vortex penetrated, with
voltage-states, superconducting versus dissipative, is not quite trivial and
depends on the relative position of the measuring field $B^\ast$ and the first
zero $B_0$ of the Fraunhofer pattern, $I_c(B_0) = 0$, with theory
\cite{Clem2010} predicting 
\begin{equation}\label{eq:B0}
   B_0 = 1.41\, \Phi_0/W^2.
\end{equation}
We address this assignment with the help of the schematic in
Fig.~\ref{Fig1}(d).  If the measuring field $B^\ast$ resides below the first
zero at $B_0$ of the Fraunhofer pattern, $B_1^\ast < B_0$, see black-dashed
line in Fig.~\ref{Fig1}(d), then the finite-voltage state has no vortex
present in the leads, while the superconducting state is the one with a vortex
trapped in the lead.  On the contrary, if $B_2^\ast > B_0$, the white-dashed
line in Fig.~\ref{Fig1}(d), then the identification is reversed, the
superconducting state is the one without vortex and the dissipative state is
the one with a vortex trapped.

Obviously, such fluctuations between different (superconducting and
dissipative) states appear only within an intermediate current range:  In the
low-current regime $I < I_c,\,I_c^{\rm v}$, the junction remains in the
superconducting state with and without the vortex present in the lead, while
in the high-current region $I > I_c,\,I_c^{\rm v}$, the junction resides in
the resistive state for both lead states---this explains the homogeneous
dark-blue and light-blue regions in Figs.~\ref{Fig2}(a) and (b). Hence vortex
fluctuations can be observed as critical current jumps in the intermediate
regime $I_c < I < I_c^{\rm v}$ (if $B_1^\ast < B_0$; if $B_2^\ast > B_0$ then
the interval is $I_c^{\rm v} < I < I_c$, see Fig.~\ref{Fig1}(d)). Furthermore,
vortex fluctuations within the leads can be observed through voltage jumps in
the dissipative state. This requires the junction dissipation to depend on
lead state as well. E.g., this is the case for a dissipative junction
\cite{Tinkham2004} with a hyperbolic onset of voltage $V \propto \sqrt{I^2 -
I_c^2}$, where a shift in $I_c$ due to the presence of a vortex changes the
voltage measured at current $I > I_c$. Indeed, it is this mechanism that
allows us to observe voltage fluctuations in the rounded $V$--$I$
characteristic at high temperatures, see Sec.~\ref{sec:TN-hT} below.  Note
that a non-dissipative junction with a jump to the normal state with fixed
resistance $R$ cannot distinguish between different lead states; similarly, a
dissipative junction probed at high currents $I \gg I_c$ away from voltage onset
cannot serve as a sensitive vortex detector. With the presence/absence of
vortices in the lead mimicking a shift of the effective magnetic field seen by
the junction, see Sec.~\ref{sec:JJ} above, the suitability of a (dissipative)
junction as a vortex sensor can be tested through measuring the field
dependence of the $V$--$I$ characteristic.  We will return to this discussion
in Sec.~\ref{sec:TN} below when associating different voltage levels in the
telegraph noise (rather then critical currents in the FP) with different lead
states.

When following the evolution of the Fraunhofer pattern in Fig.~\ref{Fig2}(a)
with increasing temperatures $T$, see Figs.~\ref{Fig2}(c)--(e), we observe the
same main features as in Fig.~\ref{Fig2}(a), a noisy region around small
values of $I_c(B)$ and a stationary second peak around $B \approx
\SI{3}{mT}$.  In addition, besides some enhanced smearing due to temperature,
we can note a change in $B$-scale, i.e., all of the pattern is squeezed
towards smaller $B$-values. In particular, the position of the first zero
$B_0$ in the Fraunhofer pattern decreases with increasing temperatures, as
shown with the fits (orange lines) of the main peak in Figs.~\ref{Fig2}(a) and
(c)--(e); here, we use the width $W$ in the expression
\begin{equation}\label{eq:FP_thin_film}
   \frac{I_c(B)}{I_c(0)} = \Big|J_0\Bigl(\frac{1.705\, B W^2}{\Phi_0}\Bigr)\Big|
\end{equation}
as a fit parameter for the Fraunhofer pattern in a weakly screening film
\cite{Clem2010}.  In Fig.~\ref{Fig2}(f), we summarize this finding with a
$B_0(T)$ plot that decreases from $B_0(T=\SI{7}{mK}) \approx \SI{2.2}{mT}$ to
$B_0(T=\SI{120}{mK}) \approx \SI{1.3}{mT}$, corresponding (via
Eq.~\eqref{eq:B0}) to an increase of the relevant width from $W_{\rm
eff}(\SI{7}{mK}) \approx \SI{1.16}{\upmu m}$ to $W_{\rm eff}(\SI{120}{mK})
\approx \SI{1.5}{\upmu m}$. The observed decrease in $B_0$ with increasing
temperature $T$ goes beyond the generic model of Ref.~\cite{Clem2010} as the
width $W$ is fixed at the geometric length of the junction. However, one may
speculate on different extensions of the model that produce such an effect,
e.g., a superfluid density that gets suppressed more strongly at the junction
with increasing temperature, forcing a larger phase gradient $\partial_y
\gamma_{\rm \scriptscriptstyle J}(y)$ along the junction, see
Ref.~\cite{Clem2010}. The critical phase drop $\pi$ along the junction
producing compensating forward and backward current densities then is reached at a
smaller field $B$ than expected on purely geometrical reasons.

Below, we will further discuss the fluctuations in the junction-voltage that
we associate with vortex dynamics in the leads.  Keeping a fixed measuring
field $B^\ast = \SI{2}{mT}$ and current $I^\ast = \SI{4}{nA}$, our association
of pristine versus vortex-penetrated states with superconducting versus
dissipative states will change depending on the position of $B_0(T)$ relative
to $B^\ast$.  In particular, it follows from Fig.~\ref{Fig2}(f) that we should
expect a changeover in the association of high and low dissipation with pristine
versus vortex-penetrated states around $T \approx\SI{40}{mK}$.

\section{Noise in Voltage}\label{sec:TN}

To study single-vortex motion, we measure time-traces, see Fig.~\ref{Fig3}, of the voltage across the Josephson junction, recorded as the output of a data acquisition card, $V_\mathrm{d}(t)$ (for setup details, see Appendix \ref{appendix_B}).
We fix the magnetic field at $B^\ast=\SI{2}{mT}$ and choose a bias current
$I^\ast=\SI{4}{nA}$ that lies within the dynamical regime above the pristine
critical current $I_c(B^\ast)$; time records over 4 hours ensure statistical
significance of our measurements.  For each temperature (between $T =
\SI{7}{mK}$ and $T = \SI{120}{mK}$), we observe voltage switchings between
distinct levels, that we associate with vortices entering or exiting the
leads. The assignment of voltage levels to corresponding vortex or no-vortex
states depends on the relative position of $B^\ast$ and $B_0(T)$ and changes
with temperature as described above; special care has to be taken in the
vicinity of $B_0$ with the rounded $V$--$I$ characteristic due to the
vanishing critical current $I_c(B_0) = 0$. Here, we describe in detail the
measurement outcomes at low- ($T=\SI{7}{mK}$) and high- ($T=\SI{100}{mK}$)
temperatures which together capture the essential features of our analytic
procedure.

\subsection{Low temperature results}\label{sec:TN-lT}

\begin{figure} \includegraphics[width=\columnwidth]{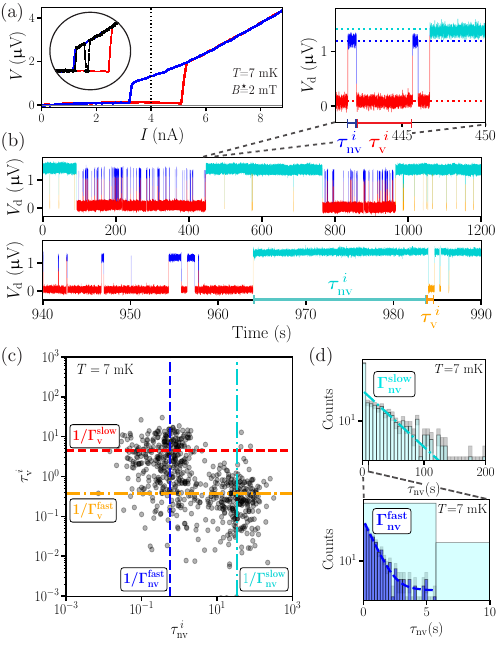}
\caption{Telegraph noise measured at low temperature $T = \SI{7}{mK}$. (a)
Voltage--current characteristic of pristine (blue) and vortex-penetrated
sample (red). The inset shows jumps within the dynamical regime where vortex
penetration can be detected. (b) Time traces of the voltage $V_\mathrm{d}(t)$ across the
junction taken at $B^\ast = \SI{2}{mT}$ and $I^\ast = \SI{4}{nA}$ (dotted line
in (a)). The telegraph noise exhibits two regimes, blue--red and cyan--orange,
with red/orange colors relating to superconducting states (switched by
fluctuating vortices) and blue/cyan colors to dissipative states (pristine or
in the presence of a distant quasi-stationary vortex). Top expansion: Note the two
dissipative levels that we associate with the pristine sample (blue) and the
presence of a distant quasi-stationary vortex (cyan). (c) Time-correlation
plot $\tau_{\rm nv}$ versus $\tau_{\rm v}$. The two regimes in (b) are
associated with separate clusters. (d) Histograms serving the determination of
slow (top) and fast (bottom) rates $\Gamma$.}
\label{Fig3} 
\end{figure}

Figure \ref{Fig3} summarizes the measurement outcome of voltage time-traces
$V_\mathrm{d}(t)$ and its analysis in the low temperature regime, here, specifically at
$T=\SI{7}{mK}$. Jumps in the $V$--$I$ trace between the superconducting and
dissipative states, see Fig.~\ref{Fig3}(a), produce time traces $V_\mathrm{d}(t)$
exhibiting telegraph-type noise as shown in Fig.~\ref{Fig3}(b), taken at fixed
bias $I^\ast=\SI{4}{nA}$. At this low temperature, we have $B^\ast < B_0$,
see Fig.~\ref{Fig2}(f), and we associate the superconducting state with the
presence of a vortex in the leads, see Fig.~\ref{Fig1}(d).  The observed
signal involves two distinct types of segments: one that is predominantly
superconducting (red) with rapid excursions to the dissipative state (blue)
and another where the junction resides mainly in the dissipative state (cyan)
with dilute switchings to the superconducting state (orange).

We aim at finding the rates for the processes appearing in the voltage traces
$V_\mathrm{d}(t)$.  In a first step, the traces are grouped into two states, dissipative
(blue or cyan) and superconducting (red or orange), and digitized using an
appropriate algorithm \cite{Yuzhelevski_2000} allowing for the identification
and extraction of the waiting times, see Appendix \ref{appendix_C} for
algorithm details. We denote by $\tau_{\rm nv}$ the time intervals with no
vortex fluctuation in the leads, i.e., blue or cyan segments.  Time intervals
with a fluctuating vortex present in the sample are denoted by $\tau_{\rm v}$
(red and orange segments).  To visualize these time intervals over the entire
4-hour measurement, we use correlation plots as shown in Fig.~\ref{Fig3}(c).
The plot is constructed as follows: the full $V_\mathrm{d}(t)$ trace is divided into
consecutive pairs of $\tau_{\rm nv}$ and $\tau_{\rm v}$ intervals. Each point
on the correlation plot then represents the $i$-th no-vortex time $\tau_{\rm
nv}$ versus the following vortex time $\tau_{\rm v}$.

\begin{figure*}
    
\includegraphics[width=1.5\columnwidth]{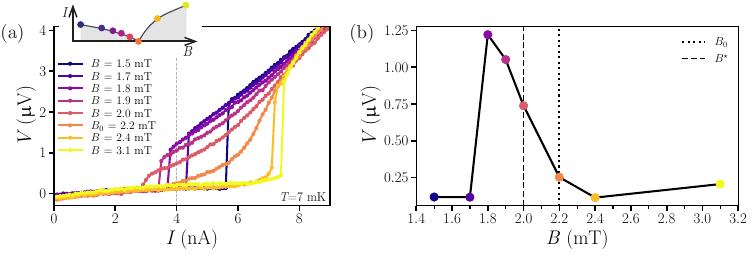}
\caption{(a) Evolution of $V$--$I$ characteristic with increasing field $B$
passing the first zero $B_0$ of the Fraunhofer pattern. The sharp critical
current $I_c(B < B_0)$ disappears on approaching $B_0$ (dark violet to pink),
giving way to a rounded characteristic at $B_0$ (orange), and returns back to
a sharp trace (bright orange and yellow) with clearly identifiable $I_c(B >
B_0)$.  (b) Dissipation at $I^\ast = \SI{4}{nA}$ versus field $B$ in the
vicinity of $B_0$. The slope in $V(B)$ is {\it negative} at the measurement
field $B^\ast = \SI{2}{mT}$ and goes through a maximum on decreasing the
field, before collapsing to zero at around $B \approx \SI{1.7}{mT}$. }
\label{Fig4}
\end{figure*}

The presence of the two types of segments described above, blue--red and
cyan--orange, is evident in the correlation plot where the data concentrate in
two clusters, see Fig.~\ref{Fig3}(c). To extract the rates for vortex-entry
and vortex-exit, the waiting times are binned into histograms following the
Freedman-Diaconis rule \cite{Freedman_Diaconis_1981}. When comparing the
timescale of the blue segments to the cyan ones (or the red to the orange
ones), we note that they differ by about one to two orders of magnitude.
Thanks to this separation of time scales, we can analyse the dissipative
intervals ($V_\mathrm{d} \neq 0$, blue and cyan segments) together by fitting
the histograms with bi-exponential distributions. The same holds for the
superconductive intervals ($V_\mathrm{d} =0$, red and orange segments). We
first fit the slower timescales (cyan and red), with the fast timescale
contribution hidden in the first bin.  Subsequently, the first bin with the
fast processes is re-binned and fitted with the sum of both exponentials, see
Fig.~\ref{Fig3}(d) and Appendix \ref{appendix_C}. In this way, a fast
($\Gamma^{\rm fast}_{\rm nv}$) and a slow rate ($\Gamma^{\rm slow}_{\rm nv}$)
are extracted for the blue and cyan segments, and similar for the
fluctuating-vortex state $\Gamma^{\rm fast}_{\rm v}$ (orange) and $\Gamma^{\rm
slow}_{\rm v}$ (red).  As shown in Fig.~\ref{Fig3}(c), the inverse of these
fitted rates accurately intersect the centres of the two clusters, supporting
our fitting approach.  The cluster centres correspond to the intersection
points ($1/\Gamma^{\rm fast}_{\rm nv}, 1/\Gamma^{\rm slow}_{\rm v}$) for the
upper left (blue--red segments), and ($1/\Gamma^{\rm fast}_{\rm v},
1/\Gamma^{\rm slow}_{\rm nv}$) for the lower right cluster (cyan--orange
segments). The same analysis is followed for all temperatures $T<\SI{40}{mK}$,
providing us with vortex-entry and -exit rates $\Gamma$ as a function of
temperature $T$.

Let us associate the various voltage levels with specific lead states.  The
two types of segments not only differ in their switching dynamics but also in
the voltage level of the blue and cyan dissipative states, see the blow-up in
Fig.~\ref{Fig3}(b), showing that the blue state dissipates less than the cyan
state. We can make use of this observation in associating the voltage levels
to different lead states, see also the discussion of the dissipative vortex
detector in Sec.~\ref{sec:FP} above.  We first trace the evolution of the
$V$--$I$ characteristic when changing $B$ across $B_0$ (as a vortex mimicks a
shift in effective $B$ seen by the junction as discussed in
Sec.~\ref{sec:JJ}), see Fig.~\ref{Fig4}(a). The sharp step marking the
critical junction current decreases on approaching $B_0\approx\SI{2.2}{mT}$,
see characteristics at $B = 1.5,~1.7,~1.8,~\SI{1.9}{mT}$ in Fig.~\ref{Fig4}(a). Around
$B_0$, the $V$--$I$ trace becomes rounded, see characteristics at
$B = 2.0,~\SI{2.2}{mT}$. Finally, a finite critical current
reappears above $B_0$, see characteristics at $B =
2.4,~\SI{3.1}{mT}$ in Fig.~\ref{Fig4}(a).  Furthermore, the resistive branch strongly
depends on the magnetic field, what will help us in associating the
voltage-levels with lead states.  Analyzing the dissipation measured at
$I^\ast = \SI{4}{nA}$, the voltage increases steeply when $I_c(B)$ drops below
$I^\ast$ and then decreases more slowly on approaching $B_0$ and beyond, see
Fig.~\ref{Fig4}(b); remarkably, the dissipation $V(B)$ is quite asymmetric
with respect to $B_0$.

At the measurement field $B^\ast = \SI{2}{mT}$, the dissipation $V(B)$ shows a
negative slope, see Fig.~\ref{Fig4}(b); as the field is decreased below
$B^\ast$, the voltage first goes through a maximum and then steeply drops to
zero. When a vortex enters the lead, the effective field seen by the junction
is reduced.  A consistent association of voltage levels and lead states in
Fig.~\ref{Fig3}(b) is obtained if we associate the state with lower
dissipation (blue in Fig.~\ref{Fig3}(b)) with the pristine lead. A distant
quasi-stationary vortex decreases the effective $B$-field by a small amount
and the junction remains in the dissipative state with a {\it larger} voltage
(see Fig.~\ref{Fig4}(b)) as compared to the pristine state---hence, we
associate the state with the higher dissipation (cyan in Fig.~\ref{Fig3}(b))
with the presence of a distant vortex. On the other hand, a (fluctuating)
vortex nearby the junction reduces the effective $B$-field quite strongly (by
about $\SI{0.4}{mT}$, as follows from the shift between orange and red lines
in the Fraunhofer pattern of Fig.~\ref{Fig2}(a)) and the dissipation drops to
zero, i.e., the fluctuating vortex pushes the junction into the
superconducting state (red or orange in Fig.~\ref{Fig3}(b)).

Summarizing, the following overall picture emerges: a distant quasi-stationary
vortex (timescale $\sim \SI{e3}{s}$) slowly modifies the energy landscape for
vortices from pristine (blue dissipation level in Fig.~\ref{Fig3}(b)) to
vortex-penetrated (cyan dissipation level).  Fast (fluctuating) vortices
(timescale $\sim \SI{1}{s}$) entering/exiting the leads closer to the junction
produce a larger reduction in the effective field seen by the junction and
switch the system into the superconducting state (red and orange).

Coming back to the discussion in Sec.~\ref{sec:AJPvortices}, we remark that
the existence of different dissipative states (blue and cyan) speaks against
an interpretation of switching events due to metastable states in a long
junction: while the superconducting junction may reside in different
metastable states with different critical currents $I_c$, one expects the
dissipative state of the junction to be unique. 

\subsection{High temperature results}\label{sec:TN-hT}
\begin{figure}
\includegraphics[width=\columnwidth]{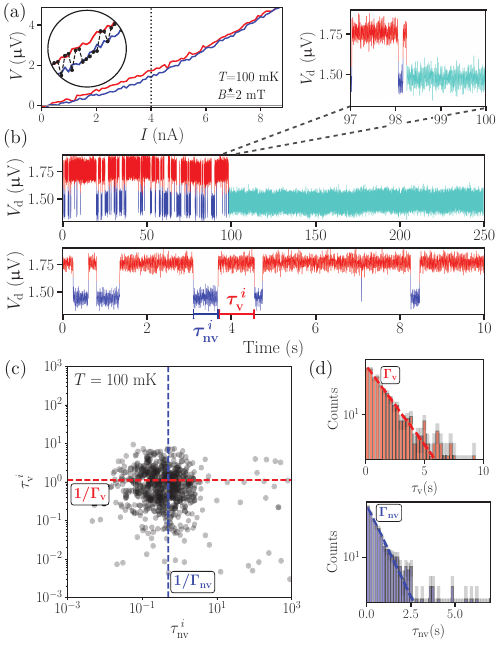}
\caption{Telegraph noise measured at high temperature $T = \SI{100}{mK}$. (a)
Voltage--current characteristic of pristine sample (blue) below the
vortex-penetrated sample (red); note the difference to Fig.~\ref{Fig3}(a)
where the pristine sample resides above the vortex-trapped one. The inset
shows jumps within the dynamical regime where vortex penetration can be
detected. (b) Time traces of the voltage $V_\mathrm{d}(t)$ across the junction taken at
$B^\ast = \SI{2}{mT}$ and $I^\ast = \SI{4}{nA}$ (dotted line in (a)). The
cyan--orange regimes present at low temperatures have turned silent (no orange
spikes due to fluctuating vortices).  (c) Time-correlation plot $\tau_{\rm nv}$
versus $\tau_{\rm v}$ with only one cluster as the cyan regime has turned
silent. (d) Histograms serving the determination of rates $\Gamma$.} 
\label{Fig5} 
\end{figure}

When going to high temperatures ($T>\SI{40}{mK}$), two modifications have to
be considered. First, the superconducting transition becomes rounded and no
sharp critical current can be identified, such that stochastic switching now
happens between two dissipative states, see Fig.~\ref{Fig5}(a). And second,
$B^\ast$ resides above $B_0$, $B^\ast = \SI{2}{mT} > B_0$, see
Fig.~\ref{Fig2}(f), such that vortex entry causes the voltage to jump upwards
to the more dissipative state, see the above explanation and schematic in
Fig.~\ref{Fig1}(d) (white dashed line at $B^\ast_2$).  The same procedure for
the analysis of the time intervals as described above is followed for $T >
\SI{40}{mK}$, with the zero-voltage level now turning dissipative.

Upon increasing temperature, the characteristic time $\tau_{\rm v}^{\rm fast}$
associated with the orange segments---during which a fluctuating vortex
resides in the lead---progressively decreases, see Fig.~\ref{Fig6}(b) with
$\Gamma^{\rm fast}_{\rm v}$ rapidly increasing in the interval $\SI{80}{mK} <
T < \SI{100}{mK}$.  Eventually, $\tau_{\rm v}^{\rm fast}$ drops below the
resolution limit of the detector (with bandwidth $f_{\rm \scriptscriptstyle
BW}\approx\SI{1.1}{kHz}$, see Appendix \ref{appendix_C}), such that the
corresponding fast dynamics can no longer be resolved. As a consequence, the
cyan(--orange) segments appear effectively silent, with no observable (orange)
voltage jumps for $T > \SI{100}{mK}$ as shown in Fig.~\ref{Fig5}(b).
Accordingly, the correlation plot now involves only a single (blue--red)
cluster, see Fig.~\ref{Fig5}(c); the scattered data to the right of the
cluster arise from the silent and hence long cyan segments.  As shown in
Fig.~\ref{Fig5}(d), the data are then fitted with a single exponential
distribution, from which the vortex rates for the blue--red segments,
$\Gamma^{\rm fast}_{\rm nv}$ and $\Gamma^{\rm slow}_{\rm v}$, are extracted.
The inverse of these rates capture well the centre of the cluster shown in
Fig.~\ref{Fig5}(c).

The extracted temperature dependences of the rates for the blue--red segments,
$\Gamma^{\rm fast}_{\rm nv}$ and $\Gamma^{\rm slow}_{\rm v}$, and cyan--orange
segments, $\Gamma^{\rm slow}_{\rm nv}$ and $\Gamma^{\rm fast}_{\rm v}$, are
shown in Figs.~\ref{Fig6}(a) and (b). Both pairs of rates show the same
behaviour: they are about constant below the temperature $T \approx
\SI{80}{mK}$ and rapidly increase at higher $T$.  Note
that at temperatures above $T > \SI{120}{mK}$, the detector resolution is
limited by thermal noise, preventing a reliable extraction of rates.
In addition, around $T \approx \SI{45}{mK}$, where $B^\ast \approx B_0$, the
signal resolution is reduced and vortex rates cannot be determined. In the
following, we briefly summarize the interpretation of the temperature
dependence of the measured rates within our vortex-framework; these findings
have been discussed before in our previous work \cite{perego2024experimental,
perego2026tunneling}.

\section{Discussion of Switching Rates}\label{sec:Dis}

\begin{figure*} \includegraphics[width=1.5\columnwidth]{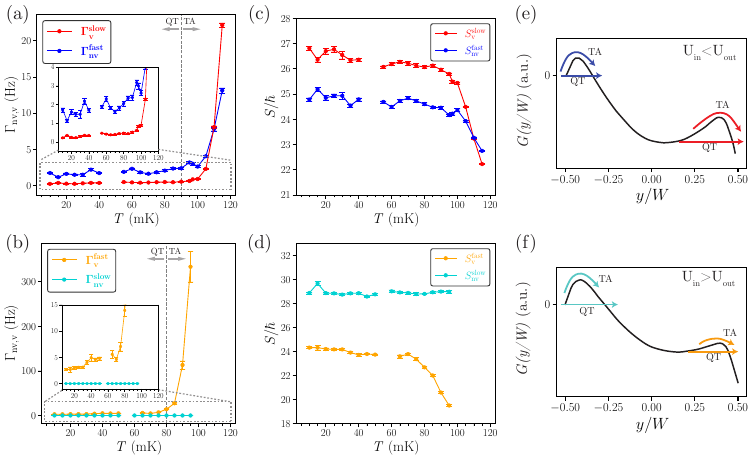}
\caption{(a) and (b) Temperature evolution of rates for vortex-entry into-
($\Gamma_{\rm nv}(T)$) and vortex-exit out ($\Gamma_{\rm v}(T)$) of the
superconducting leads. Steep rises at high temperatures are associated with
thermally activated processes over barriers. Fits with an Arrhenius law
provide barriers $U/k_{\rm\scriptscriptstyle B}$ in the Kelvin range. The
saturation of rates at low temperatures hints to macroscopic quantum tunneling
of vortices through barriers. The cyan trace stops when rapid vortex
fluctuations (orange) become unobservable due to the finite detector bandwidth
of $\SI{1.1}{kHz}$. (c) and (d) Rates $\Gamma$ transformed to dimensionless
actions $S/\hbar = \ln (\nu_0/\Gamma)$ (with $\nu_0 = \SI{2 e11}{Hz}$),
saturating between 24 and 27 at low temperatures and dropping at high
temperatures as the tunneling process gets assisted by thermal activation. (e)
and (f) Illustration of free-energy landscapes $G(y/W)$ with edge barriers for
vortex entry and vortex exit.}
\label{Fig6} 
\end{figure*}

We interpret the blue--red segments and the cyan--orange segments as
originating from rapid vortex fluctuations in the leads of our JJ (red and
orange states), once in the pristine sample (blue) and the other time with a
distant quasi-stationary vortex trapped (cyan) in one of the leads.
Figure~\ref{Fig6}(a) shows the rates associated with the vortex fluctuation in
the pristine sample as a function of temperature: $\Gamma^{\rm fast}_{\rm nv}$
corresponds to the rate of vortex entry, while $\Gamma^{\rm slow}_{\rm v}$
corresponds to the rate of vortex exit. The same interpretation applies to the
rates shown in Fig.~\ref{Fig6}(b), which describe the dynamics of the vortex
fluctuation in the presence of an additional distant quasi-stationary vortex.
In this case, $\Gamma^{\rm slow}_{\rm nv}$ corresponds to the vortex entry
rate, while $\Gamma^{\rm fast}_{\rm v}$ describes the vortex exit rate.

The temperature dependence of these rates suggests that we deal with a thermal
activation process at high temperatures above $T \approx \SI{100}{mK}$ as
described by rates of the form
\begin{align} \label{eq:th}
    \Gamma_{\rm nv,v}(T) = \nu_0 \exp \left( -U_{\rm nv,v}/
    k_{\rm\scriptscriptstyle B} T \right),
\end{align}
with $U_{\rm nv,v}$ the energy barrier, $\nu_0$ the attempt frequency, and
$k_{\rm\scriptscriptstyle B}$ is the Boltzmann constant. Fitting the
high-temperature part of the rates with such an Arrhenius law, we can extract
$\nu_0$ in the range $\sim \SI{e11}{Hz}$ (cf.\ \cite{Malozemoff1990,
Koshelev1990}) and barriers $U/k_{\rm\scriptscriptstyle B}$ of a few Kelvin;
e.g., from the vortex dwell time (red state) in the pristine sample and fully
developed thermal activation, we find the exit barrier $U_{\rm v}^{\rm
slow}/k_{\rm\scriptscriptstyle B} \approx \SI{2.6}{K}$ and $\nu_0 \sim \SI{2.0
e11}{Hz}$. For the lead penetrated by a distant quasi-stationary vortex (cyan
state), the fluctuating vortex (orange) has to overcome a smaller exit barrier
$U_{\rm v}^{\rm fast}/k_{\rm \scriptscriptstyle B}\approx \SI{2.0}{K}$ with
$\nu_0\approx \SI{7.2 e11}{Hz}$. On the other hand, for vortex entry into the
sample the data does not show (cyan) or only hardly shows (blue) a fully
developed exponential rise.

At low temperatures the rates saturate, suggesting a crossover to a regime
where vortices tunnel across barriers rather than jumping over
\cite{caldeira1983quantum, larkin1984quantum, glazman1984}. The rates then
give access to the action $S$ governing the tunneling process,
\begin{align} \label{eq:qu}
    \Gamma_{\rm nv,v} =\nu_0'
    \exp \left( -{S_{\rm nv,v}}/{\hbar} \right),
\end{align}
where the thermal exponent $U/k_{\rm\scriptscriptstyle B}T$ in Equation~\eqref{eq:th}
is replaced by the dimensionless quantum action $S/\hbar$ (here $S_{\rm nv,v}$
denote actions for vortex entry and exit). Figures~\ref{Fig6}(c) and (d) show
the rates plotted in terms of $S_{\rm nv,v}/\hbar = \ln(\nu_0'/\Gamma_{\rm
nv,v})$ with $\nu_0' = \nu_0$. The data is approximately temperature
independent at low temperatures, saturating to values $S/\hbar \approx 24$ --
$27$; the slight decrease upon increasing $T$ indicates that thermal
fluctuations progressively assist the tunneling process. This regime is
followed by a rapid decay at higher temperatures with a crossover temperature
$T_0$ defined by the matching of the exponents in Eqs.~\eqref{eq:th} and
\eqref{eq:qu}, $S(T_0)/\hbar \approx U/k_{\rm\scriptscriptstyle B} T_0$.

Let us verify that the obtained barriers $U/k_{\rm\scriptscriptstyle B}$ in
the Kelvin- and actions $S/\hbar$ in the few-tens- range make physical sense
in the situation at hand. The free-energy landscape $G(y)$ for a vortex
crossing a thin-film strip in the weak-screening limit has been calculated in
Ref.~\cite{Stejic1994}. The free energy rises steeply at the film edges due to
the Bean-Livingston surface barrier \cite{BeanLivingston1964}. A finite
magnetic field $B \approx \mu_0 H$ produces a minimum in the film center,
while the transport current $I$ tilts the landscape. In the tilted landscape,
a vortex cannot tunnel in the reverse direction, what lets us associate the
observed fluctuations at low temperatures with vortices traversing the film.
Barrier heights $U$ are given by the vortex energy $\varepsilon_0 d$, while
barrier widths are a fraction of $W$, see Figs.~\ref{Fig6}(e) and (f) for
typical shapes of the free-energy landscape.

Comparing our experimental result $U/k_{\rm \scriptscriptstyle B} \approx
\SI{2.6}{K}$ with the theoretical finding that $U \approx \varepsilon_0 d$, we
find \cite{perego2024experimental} a superfluid density $\rho_s/k_{\rm
\scriptscriptstyle B} \approx \SI{0.8}{K}$ and a value $\lambda_{\rm
\scriptscriptstyle L} \approx \SI{2.7}{\upmu m}$ for the London penetration
length, in agreement with other results \cite{Kim2024,Oliver2024}. The action
$S$ for a dissipative tunneling process is determined by the tunneling
distance $\sim q_b$ and the film resistance $R^{\scriptscriptstyle \square} =
\rho_n/d$, $S \approx \hbar (\pi/8) (q_b/\xi)^2 (R_{\rm\scriptscriptstyle K}/
R^{\scriptscriptstyle \square})$, where $q_b$ is the distance between the
minimum ($G=0$) and maximum ($G = U$) of the tunneling barrier as modelled by
a cubic parabola \cite{larkin1984quantum}.  Here, $R_{\rm\scriptscriptstyle K}
= h/e^2 \approx \SI{25.8}{k\Omega}$ is the von-Klitzing constant and $\xi$ is
the coherence length. With $R^{\scriptscriptstyle\square} \approx
\SI{2}{k\Omega}$ and a coherence length \cite{perego2024experimental} $\xi
\approx \SI{40}{nm}$, we find a tunneling distance $(3/2) q_b \approx W/8
\approx 3.5 \, \xi$ that compares well with the typical barrier thickness in
the free-energy landscape $G(y)$.

\section{Summary}\label{sec:Sum}

In summary, we have demonstrated the versatility of our Josephson junction
sensor as a Pearl-vortex detector in a thin film. Our transport-based method
enables the detection of individual vortices, providing a framework to trace
and characterize single-vortex motion as a function of temperature and
magnetic field. The observed Pearl-vortex dynamics exhibits a complex
structure that we have attributed to the joint action of two types of
vortices: those rapidly penetrating the superconducting leads in the junction
vicinity and other quasi-stationary vortices trapped further away from the
junction. The latter control the former through a change in the pinning
landscape, specifically, their entry and exit barriers. The fast vortex
fluctuations and slow control vortices produce distinct segments in the voltage
traces.  We could disentangle the various processes and find the rates for
vortex entry- $\Gamma_{\rm nv}$ and exit $\Gamma_{\rm v}$ events in
pristine and vortex-penetrated lead states.  The temperature dependence of
these rates hints to creep-type vortex motion, thermal activation over
barriers at high temperatures and quantum tunneling through barriers at low
temperatures.  We have extracted key phenomenological parameters, including
the edge barrier $U$, the tunneling action $S$, and the attempt frequency
$\nu_0$.  We conclude that our single-vortex Josephson sensor provides us with
valuable insights into the behavior of Pearl-vortices in twisted-layer
graphene and opens the possibility to integrate them into the functionality of
future superconducting electronics \cite{Golod2022}.

\section*{Data availability}
The data supporting the findings of this study, together with the code for
plotting the figures, is available online through the ETH Research Collection
at \url{https://doi.org/10.3929/ethz-c-000802057}.

\begin{acknowledgments}
We thank Peter M\"{a}rki and the staff of the ETH cleanroom facility FIRST for
technical support. We acknowledge fruitful discussions with Vladimir Kogan.
Financial support was provided by the European Graphene Flagship Core3
Project, H2020 European Research Council (ERC) Synergy Grant under Grant
Agreement 951541, the European Union’s Horizon 2020 research and innovation
program under grant agreement number 862660/QUANTUM E LEAPS, the European
Innovation Council under grant agreement number 101046231/FantastiCOF, the EU
Cost Action CA21144 (SUPERQUMAP).  K.W. and T.T. acknowledge support from the
JSPS KAKENHI (Grant Numbers 21H05233 and 23H02052) and the World Premier
International Research Center Initiative (WPI), MEXT, Japan.C.G.A. acknowledges support from the Heidi Ras foundation via an ETH Quantum Center fellowship. 

\end{acknowledgments}

\subsection*{Author contributions}
M.P. fabricated the device. T.T. and K.W. supplied the hBN crystals. M.P.,
P.K. and C.G.A. performed the measurements. M.P. and P.K. analysed the data.
V.G. and G.B. developed the theoretical model. M.P. and G.B. wrote the
manuscript, and all authors were involved in the reviewing process. M.P.,
P.K., C.G.A. and A.M.T. discussed the data. M.P., V.G., G.B., K.E. and T.I.
conceived and designed the experiment. T.I. and K.E. supervised the experimental work.

\appendix
\renewcommand{\thefigure}{A\arabic{figure}}
\setcounter{figure}{0}
\section{Fabrication and tuning details}
\label{appendix_A}
The details for the tuning, fabrication, capacitance model and angle estimates
can be found in Ref.~\cite{perego2024experimental} and its Supplementary
Material. There, the main focus is on the \textit{strong-leads} tuning with
the leads carrier density and displacement field fixed to $n_\textnormal{l} =
\SI{4.2e12}{cm^{-2}}$ and $D_\textnormal{l}/\epsilon_0 = \SI{-0.3}{V/nm}$.
The \textit{weak-leads} tuning discussed in the present work, instead, has the
leads tuned to $n_\textnormal{l} = \SI{4.8e12}{cm^{-2}}$ and
$D_\textnormal{l}/\epsilon_0 = \SI{-0.37}{V/nm}$.  In all measurements, the
junction region is tuned such that the junction carrier density and
displacement field are $n_\textnormal{j} = \SI{6.2e12}{cm^{-2}}$ and
$D_\textnormal{j}/\epsilon_0 = \SI{-0.5}{V/nm}$.

\section{Measurement setup}
\label{appendix_B}
\begin{figure*}
    \centering
    \includegraphics[width=2\columnwidth]{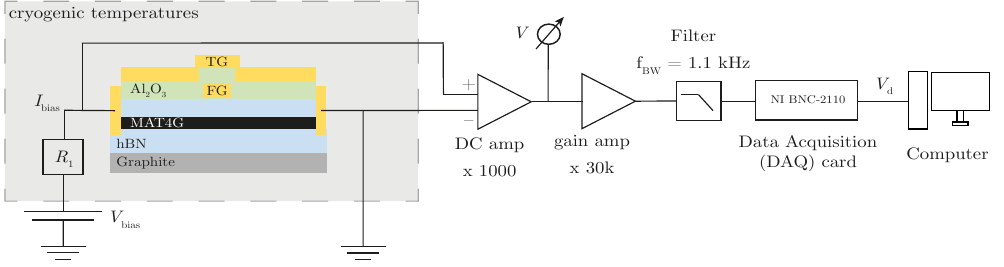}
    \caption{Measurement setup. A DC voltage source provides a bias current, $I_{\mathrm{bias}}$, to the sample. The voltage drop across the junction is amplified by a DC differential amplifier, yielding an intermediate voltage $V$. This signal is further amplified and low-pass filtered before being input to a Data Acquisition (DAQ) card. The final voltage, $V_{\mathrm{d}}$, is then recorded by a computer.} \label{fig: Msmt_Sketch}
\end{figure*}

All measurements were performed in a dilution refrigerator at a base
temperature of 7 mK, see setup sketch in Fig.~\ref{fig: Msmt_Sketch}. We employed a current-biased $I_\mathrm{bias}$, two-terminal configuration,
subtracting contact resistances during post-processing. Bias was generated using a home-built d.c\ source in series with a $R_1 = \SI{100}{M\Omega}$ resistor. The resulting voltage drop was amplified ($\times 1000$) via a
low-noise d.c.\ amplifier (see \cite{marki2017temperature}) and recorded by a HP 3441A multimeter, voltage output $V$. This voltage output was used for all the measurements in this work, except the vortex time trace for statistical analysis. For vortex time trace measurements, the voltage was
further amplified ($\times 30000$), low-pass filtered at $\SI{1.1}{kHz}$ and sampled at $\SI{20}{kHz}$ using an NI BNC-2110 Data Acquistion (DAQ) card. The voltage output from the data card $V_\mathrm{d}$ was then recorded by a computer. These setup parameters were optimized for signal-to-noise ratio. Gate voltages for the top, finger and bottom gates were controlled by home-built low-noise d.c.\ voltage sources.

\section{Data analysis of voltage traces $V_\mathrm{d}(t)$}
\label{appendix_C}
\subsection{Post-processing}

Voltage traces $V_\mathrm{d}(t)$ are first filtered using a notch filter at $f_{\rm
cable} = \SI{661\pm20}{Hz}$ to remove cabling noise. Since the relevant signal
information is confined below $\SI{100}{Hz}$, we apply a second-order digital
low-pass filter with a cutoff of $f_{\rm LP} = \SI{200}{Hz}$. The data are
then resampled at $f_{\rm resample} = \SI{3}{kHz}$ to optimize computational
efficiency; this preserves the time-domain signal integrity as the Nyquist
frequency ($\SI{1.5}{kHz}$) remains well above the filtered bandwidth
\cite{Gustavsson_PHD}. Discrete switching events are extracted by digitizing
the signal using the Yuzhelevski algorithm \cite{Yuzhelevski_2000}. Five
iterations of this process \cite{Gustavsson_PHD} yield a clean two-state
trace, from which we determine the waiting times for both voltage states.

\subsection{Fitting procedure}

Waiting time statistics are analysed by binning the data according to the
Freedman-Diaconis rule \cite{Freedman_Diaconis_1981}, with manual adjustments
made for cases of severe under- or over-binning. As shown in
Fig.~\ref{Fig3}(d), the distributions exhibit a bi-exponential dependence
rather than a single decay, characterized by a fast timescale in the initial
bins and a slow timescale in the tail. We model this behaviour using
\begin{equation} \label{eq:bi-exponential} 
   f_i(t) \propto A_i^{\rm fast} e^{-\Gamma_i^{\rm fast} t} 
                + A_i^{\rm slow} e^{-\Gamma_i^{\rm slow} t},
\end{equation} 
where $i \in \{ \rm vortex, \rm no-vortex\}$. To extract the fit parameters,
we first fit the slow timescale where the fast contribution is negligible.
Then, the first bins are re-binned and fit with the full bi-exponential sum.
Parameter errors are derived from least-square minimization and propagated
linearly.

\subsection{Finite-bandwidth correction}
Extracted rates are bandwidth-corrected to account for the systematic
underestimation caused by finite detector response times \cite{Naaman_2006}.
In a two-level system, the detector may miss rapid sequential transitions
between the two true states, $A$ and $B$. If a fluctuation (e.g., $A \to B \to
A$) occurs faster than the detection rate $\Gamma_{\rm det}$, the system fails
to register the brief excursion to $B$. Consequently, the detector erroneously
records a single, persistent "observed" state ($A^\star$) instead of the true
underlying dynamics. We apply the correction from \cite{Naaman_2006},
\begin{align}
   \Gamma_v &= \Gamma_{v^\star}\frac{\Gamma_{\rm det}}
              {\Gamma_{\rm det}-\Gamma_{v^\star}-\Gamma_{nv^\star}}, \\
   \Gamma_{nv} &= \Gamma_{nv^\star}\frac{\Gamma_{\rm det}}
                 {\Gamma_{\rm det}-\Gamma_{v^\star}-\Gamma_{nv^\star}},
\end{align}
where $\star$ denotes the uncorrected rates obtained in the fitting procedure.
This adjustment recovers the events lost at short waiting times, ensuring the
decay rates accurately reflect the intrinsic vortex dynamics.

To use the above formulae, we must determine the detection rate $\Gamma_{\rm
det}$. We analyse the transition time of voltage switching events, which
exhibit a non-instantaneous delay due to the measurement bandwidth. By
averaging hundreds of these events and fitting the delay to an exponential
decay $\propto \exp(-\Gamma_{\rm det} t)$, we extract $\Gamma_{\rm det}$.
Across multiple temperatures, we find $\Gamma_{\rm det} = \SI{1098 \pm
93}{Hz}$, consistent with the setup's expected $\SI{1100}{Hz}$ limit.

\bibliography{Bibliography}%

\end{document}